Structure and energetics of carbon-related defects in SiC (0001)/SiO$_2$ systems revealed by first-principles calculations: Defects in SiC, SiO$_2$, and just at their interface

Takuma Kobayashi[1] and Yu-ichiro Matsushita[1]

[1]Laboratory for Materials and Structures, Institute of Innovative Research, Tokyo Institute of Technology, Yokohama 226-8503, Japan

We report first-principles calculations that reveal the atomic forms, stability, and energy levels of carbon-related defects in SiC (0001)/SiO$_2$ systems. We clarify the stable position (SiC side, SiO$_2$ side, or just at the SiC/SiO$_2$ interface) of defects depending on the oxidation environment. Under an O-rich condition, the di-carbon antisite ((C$_2$)$_{Si}$) in the SiC side is stable and critical for *n*-channel MOSFETs, whereas the di-carbon defect (Si-C-C-Si) at the interface becomes critical under an O-poor condition. Our results suggest that the oxidation of SiC under a high-temperature O-poor condition is favorable in reducing the defects, in consistent with recent experimental reports.

Silicon carbide (SiC) has attracted increasing attention as a suitable material for next-generation power devices, owing to its superior physical properties (e.g., wide bandgap, high critical electric field, and high thermal conductivity).[1,2] One of the unique advantages of SiC is that it can be thermally-oxidized to form an insulator, silicon dioxide ($SiO_2$), on its surface. Thus, metal-oxide-semiconductor field-effect transistors (MOSFETs) have been recognized as a strong candidate for power switching devices.[2]

Despite their great potential, SiC MOSFETs have suffered from the low channel mobility, which is certainly due to the high interface state density ($D_{IT}$) of $SiC/SiO_2$ systems (~ $10^{13}$ eV$^{-1}$cm$^{-2}$) near the conduction band edge ($E_C$) of SiC.[2-7] Due to this high $D_{IT}$, which is thousand-fold higher than that in silicon (Si)/$SiO_2$ systems (~ $10^{10}$ eV$^{-1}$cm$^{-2}$),[8] the field-effect mobility of SiC MOSFETs is mainly limited by the carrier trapping effect by the interface states, even after proper interface nitridation.[7]

Although the physical origin of the interface states is not identified yet, carbon-related defects have widely been believed as the major origin of interface states. In fact, a study based on internal photoemission (IPE) suggested the existence of carbon clusters near the $SiC/SiO_2$ interface.[9] A recent report based on secondary ion mass spectrometry (SIMS) also suggested the evidence of carbon-related defects and discussed the correlation between the interface quality and the excess carbon density.[10]

Theoretical calculations also focused on the carbon-related defects and suggested various forms of carbon defects so far.[11-20] As a few examples, di-carbon antisite (($C_2$)$_{Si}$)[11] in the SiC side, $Si_2$-C-O[12] and $Si_2$-C=C-$Si_2$[13-15] structures in the $SiO_2$ side, Si-(CO)-Si bridge[11] and ethylene-like Si-(CO)$_2$-Si[16] structure just at the interface have been suggested. However, very few of the reports systematically deal with the defects in the entire area around the interface (i.e., SiC side, $SiO_2$ side, and just at the $SiC/SiO_2$ interface).

In this report, we provide systematic first-principles calculations that reveal the possible atomic forms of carbon-related defects in the SiC side, $SiO_2$ side, and just at the $SiC/SiO_2$ interface. We compare the formation energies of the defects under different chemical potential conditions to discuss the defect stability depending on the oxidation environment. We also calculate the localized levels induced by the defects for those which are stable.

All of the calculations in this study were performed on the basis of density functional theory (DFT)[21,22] by using the Vienna ab initio simulation package (VASP).[23,24] Both the structural optimization and one-electron energy level calculations were performed by using the Heyd-Scuseria-Ernzerhof (HSE06) hybrid functional[25-27] to reproduce the experimental bandgap of SiC well. The projector augmented wave (PAW) method[28] as implemented in the VASP code was applied in the calculations. The cutoff energy was 400 eV in the plane-wave-basis set, and the structural optimization was performed until the remaining forces of the structures became less than 40 meVÅ$^{-1}$. A single $k$-point (the Γ point) was sampled during the calculations.

We used a 128-atom 4H-SiC supercell (a 4×4×1 supercell), a 72-atom $α$-quartz-$SiO_2$ supercell, and a 128-atom 4H-SiC surface slab as basic structures to investigate the carbon defects in SiC, $SiO_2$, and just at their interface, respectively (see supplementary data in detail). We considered 114 patterns of initial structures (SiC-side: 23 patterns, $SiO_2$-side: 12 patterns, SiC (0001)/$SiO_2$ interface: 79 patterns) to investigate a variety of mono- and di-carbon defects, which is also explained in detail in the supplementary data. The formation energy, $E_F$, of defects was calculated by

$$E_\mathrm{F} = E_\mathrm{D} - E_0 - \sum_i N_i \mu_i, \qquad (1)$$

where $E_D$ and $E_0$ are the total energy of the simulation cell with and without the defect, respectively. $N_i$ and $\mu_i$ are number of the removed ($n_i < 0$) or added ($n_i > 0$) $i$-type atom and its chemical potential, respectively. We supposed four extreme situations during the oxidation of SiC, where the chemical potentials of the atomic species were calculated as

$$\mu_{\text{Si}} = E(\text{SiO}_2) - 2\mu_\text{O}, \ \mu_\text{C} = E(\text{SiC}) - \mu_{\text{Si}}, \ \mu_\text{O} = \frac{E(\text{O}_2)}{2} \quad \text{(C-rich, O-rich)}, \tag{2}$$

$$\mu_{\text{Si}} = E(\text{Si}), \ \mu_\text{C} = E(\text{CO}_2) - 2\mu_\text{O}, \ \mu_\text{O} = \frac{E(\text{SiO}_2) - \mu_{\text{Si}}}{2} \quad \text{(C-rich, O-poor)}, \tag{3}$$

$$\mu_{\text{Si}} = E(\text{SiO}_2) - 2\mu_\text{O}, \ \mu_\text{C} = E(\text{CO}_2) - 2\mu_\text{O}, \ \mu_\text{O} = \frac{E(\text{O}_2)}{2} \quad \text{(C-poor, O-rich)}, \tag{4}$$

$$\mu_{\text{Si}} = E(\text{Si}), \ \mu_\text{C} = E(\text{SiC}) - \mu_{\text{Si}}, \ \mu_\text{O} = \frac{E(\text{SiO}_2) - \mu_{\text{Si}}}{2} \quad \text{(C-poor, O-poor)}. \tag{5}$$

Here, $E(\text{Si})$, $E(\text{CO}_2)$, $E(\text{SiO}_2)$, $E(\text{O}_2)$, and $E(\text{SiC})$ are the total energies of crystal silicon (Si), a carbon dioxide ($CO_2$) molecule, α-quartz $SiO_2$, an oxygen ($O_2$) molecule, and 4H-SiC per formula unit, respectively. We determined the relationships between the chemical potentials (Eqs. (2)-(5)) as follows; First, we assumed chemical equilibrium among Si, $O_2$, and $SiO_2$ during the oxidation of SiC and considered the O-rich or O-poor condition, where $\mu_{\text{Si}}$ and $\mu_\text{O}$ are determined as $\mu_{\text{Si}} = E(\text{SiO}_2) - 2\mu_\text{O}$ and $\mu_\text{O} = E(\text{O}_2)/2$ or $\mu_{\text{Si}} = E(\text{Si})$ and $\mu_\text{O} = (E(\text{SiO}_2) - \mu_{\text{Si}})/2$. Then, we supposed equilibrium with either SiC or $CO_2$, and $\mu_\text{C}$ was determined as $\mu_\text{C} = E(\text{SiC}) - \mu_{\text{Si}}$ or $\mu_\text{C} = E(\text{CO}_2) - 2\mu_\text{O}$, resulting in Eqs. (2)-(5). Since we focused on the C-related defects, we considered the C-rich limits (Eqs. (2) and (3)) in this study. As we approximated the $SiC/SiO_2$ interface with an H-terminated SiC surface slab (see supplementary data), the chemical potential of an H atom is also needed to calculate the formation energy (Eq. (1)), which was calculated as

$$\mu_\text{H} = \frac{E(\text{SiH}_4) - \mu_{\text{Si}}}{4}, \tag{6}$$

irrespective of the condition (Eq. (2) or (3)). Here, $E(\text{SiH}_4)$ is the total energy of a silane ($SiH_4$) molecule per formula unit. Note that the H atoms were introduced in order to eliminate the fictitious dangling bonds in the calculation model (see supplementary data). We adopted $E(SiH_4)$ in calculating $\mu_H$, since most of the H atoms were used to terminate the fictitious Si-dangling bonds. We obtained the temperature dependence of enthalpy, $H(T)$, and entropy, $S(T)$, of gas molecules, $O_2$, CO, $CO_2$, and $SiH_4$ at standard state pressure ($p^0 = 0.1$ MPa) from thermochemical tables[29] to correct their energies at finite temperature.

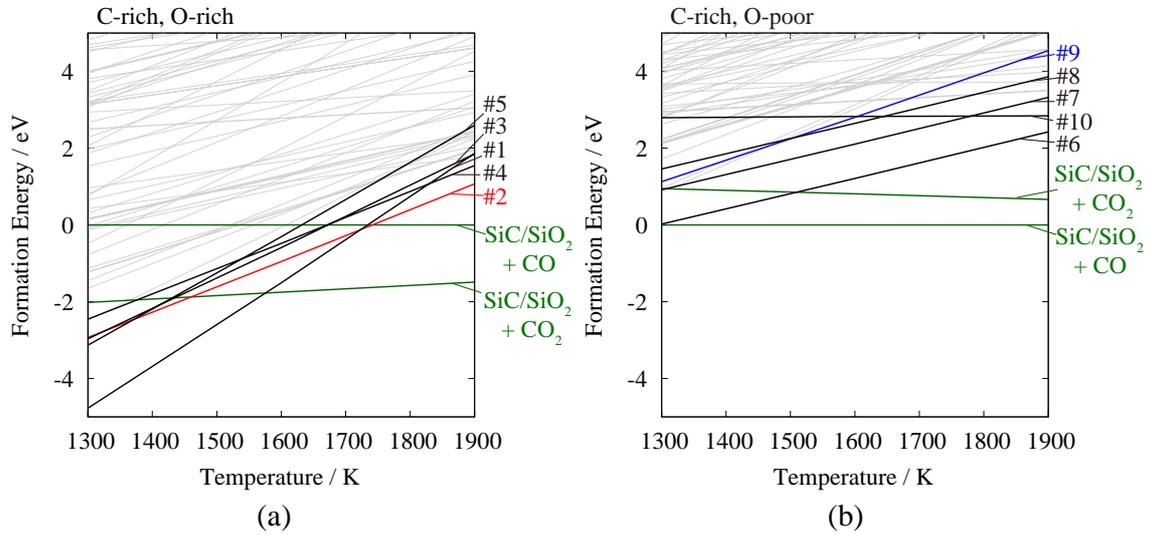

Fig. 1. Temperature dependence of defect formation energies in comparison with the energies of gaseous CO and $CO_2$ molecules at the (a) C-rich, O-rich and (b) C-rich, O-poor limits. The zeros of the formation energies are set at the energy of a CO molecule. The structure of the defects (#1 - #10) is depicted in Fig. 2. Lines in dark contrast (red, blue, and black) indicate the defects stable at 1600 K, which is about the typical experimental temperature of SiC oxidation. The red, blue, and black solid lines indicate the defects in SiC side, $SiO_2$ side, and just at the $SiC/SiO_2$ interface, respectively. Green lines indicate the gaseous CO and $CO_2$ molecules.

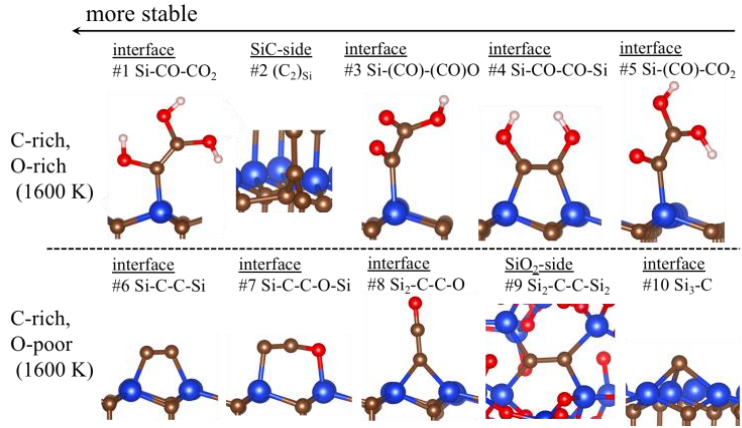

Fig. 2. Structure of the defects stable at 1600 K, which is about the typical experimental temperature of SiC oxidation. The structure is listed in order of stability, depending on the oxidation environment (the C-rich, O-rich and C-rich, O-poor limits). The blue, brown, red, and small white balls are the Si, C, O, and H atoms, respectively. Note that the H atoms are fictitious, and in the realistic situation, defects are supposed to be attached to either Si or O atoms of amorphous $SiO_2$ instead of H atoms.

Figures 1 (a) and (b) show the temperature dependence of formation energies for the carbon defects in SiC (0001)/$SiO_2$ systems. Here, the energies of gaseous CO and $CO_2$ molecules are also shown for comparison. The structure of the carbon defects is depicted in Fig. 2. At the O-rich limit (Fig. 1 (a)), the formation energies of defects #1-#5 are quite low at 1600 K, which is about the typical experimental temperature of SiC oxidation, and even comparable with the energies of gaseous CO and $CO_2$ molecules. Such a result indicates that, during the oxidation of SiC under a condition close to the O-rich limit, a significant portion of C atoms will favor staying around the interface to create defects than to be ejected as CO or $CO_2$ molecules. At the O-poor limit (Fig. 1 (b)), in contrast, the formation energies of the defects are relatively high compared to the energies of CO and $CO_2$ molecules. Therefore, in the viewpoint of static energetics, oxidation under the O-poor condition is more favorable than that under the O-rich condition in reducing the carbon defects at a SiC/$SiO_2$ interface.

When looking at the temperature dependences, CO and $CO_2$ molecules become more stable compared to the carbon defects at higher temperatures, suggesting that oxidation at higher temperatures is also favorable. In fact, a recent experimental study proved that high-temperature oxidation at 1450°C (~ 1720 K) combined with rapid cooling reduces the defects at a SiC (0001)/$SiO_2$ interface.[30] Also, post-oxidation annealing in very-low oxygen partial pressure argon (Ar) ambient at 1500°C (~ 1770 K), of which condition is considered to be close to the O-poor limit (Fig. 1 (b)), reduces defects at a SiC (0001)/$SiO_2$ interface.[31] Such results can be qualitatively explained by our calculation results.

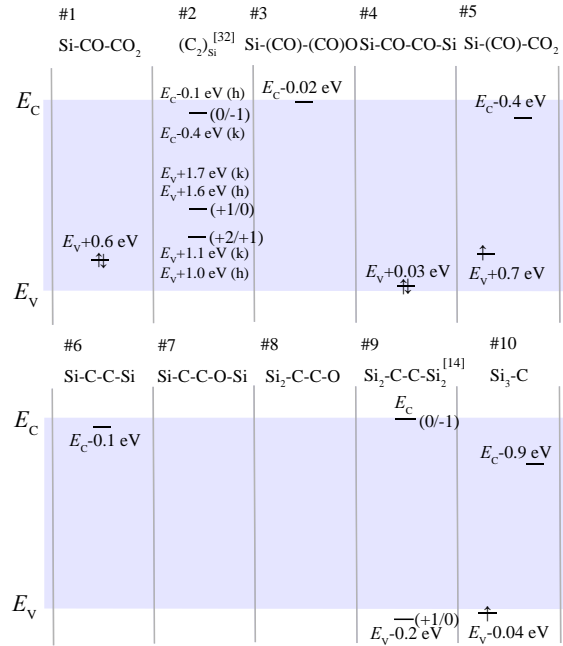

Fig. 3. Calculated electron energy levels for the defects depicted in Fig. 2. $E_C$ and $E_V$ are the conduction and valence band edges of 4H-SiC. Note that the energy levels for interface defects (#1, #3- #8, and #10) may contain an error of < 0.4 eV since the calculated bandgap (3.46 - 3.60 eV) is overestimated compared to the experiment one (3.26 eV[2]) at RT). The overestimation and variation of the calculated bandgap are attributable to the application of the slab model (see supplementary data) and to the hybridization of the defect states with the conduction and valence band states. The charge transition levels for SiC-side (#2) and $SiO_2$-side (#9) defects are taken from Refs. 32 and 14, respectively. The results of both h- and k-site defects are shown for the SiC-side defect (#2).

Figure 3 shows the energy levels of the carbon defects stable under either the O-rich or O-poor condition at 1600 K. In Fig. 2, we see that the defects with O atoms tend to appear as stable forms in the O-rich condition. Under the O-rich condition, the di-carbon antisite in the SiC side (($C_2$)$_{Si}$ (#2)) is stable and especially critical for *n*-channel MOSFETs, since it creates a defect level very near $E_C$ (about $E_C - 0.1$ eV for h-site and $E_C - 0.4$ eV for k-site defect: Fig. 3).[32] Under the O-rich condition (Eq. (2)), the chemical potential of Si is low (Si-poor) and that of C is high (C-rich), which facilitates the creation of ($C_2$)$_{Si}$ by lowering the formation energies of silicon vacancy ($V_{Si}$) and carbon clusters at the same time. The existence of ($C_2$)$_{Si}$ near the SiC/$SiO_2$ interface is also predicted in Ref. 11. Many types of interface carbon-dimer defects, such as Si-CO-$CO_2$ (#1), Si-CO-CO-Si (#4), and Si-(CO)-$CO_2$ (#5), are stable under the O-rich condition and create defect levels relatively close to the valence band edge ($E_V$) of SiC (Fig. 3). Thus, they are critical for *p*-channel MOSFETs. For the O-poor condition, the interface carbon-dimer defect (Si-C-C-Si, #6) is most stable and critical for *n*-channel MOSFETs, since it creates a defect level very near $E_C$ (about $E_C - 0.1$ eV: Fig. 3). For *p*-channel MOSFETs, the O-poor condition seems to be quite favorable and there are few stable defects which create defect levels near $E_V$ inside the bandgap (Fig. 3). As explained in the caption of Fig. 3, however, the calculated energy levels of interface defects (#1, #3- #8, and #10) may contain an error of < 0.4 eV. In such a case, the $Si_3$-C (#10) defect may create a defect level inside the bandgap near $E_V$ and hinder electrical characteristics of *p*-channel MOSFETs. However, since the $Si_3$-C (#10) defect owns a C-dangling bond, the defect level would be relatively easily passivated by hydrogen annealing.

In summary, we investigated the possible atomic forms of carbon-related defects in SiC side, $SiO_2$ side, and just at the SiC (0001)/$SiO_2$ interface. Our results indicate that oxidation of SiC under an O-rich condition, especially at low temperatures, results in the creation of critical defects both for $n$- and $p$-channel MOSFETs. For instance, the di-carbon antisite in the SiC side (($C_2$)$_{Si}$ (#2)) is stable under the O-rich condition and creates a defect level near $E_C$, being crucial for $n$-channel MOSFETs. As critical defects for $p$-channel MOSFETs, many types of interface carbon-dimer defects, such as Si-CO-$CO_2$ (#1), Si-CO-CO-Si (#4), and Si-(CO)-$CO_2$ (#5), are stable under the O-rich condition and they create defect levels relatively close to $E_V$. Thus, in the viewpoint of static energetics, we suggest the oxidation of SiC at a high-temperature O-poor condition in order to reduce the carbon defects. Even at such a condition, however, a certain number of carbon-dimer defects (Si-C-C-Si, #6) may remain just at the interface and create defect levels very near $E_C$. In realizing high-performance $n$-channel MOSFETs, this defect should be somehow further reduced.

Acknowledgments

Computations were performed mainly at the Center for Computational Science, University of Tsukuba, and the Supercomputer Center at the Institute for Solid State Physics, The University of Tokyo. The authors acknowledge the support from JSPS Grant-in-Aid for Scientific Research (A) (Grant Nos. 18H03770 and 18H03873).

Structure and energetics of carbon-related defects in SiC (0001)/$SiO_2$ systems revealed by first-principles calculations: Defects in SiC, $SiO_2$, and just at their interface

Takuma Kobayashi[1] and Yu-ichiro Matsushita[1]

[1]Laboratory for Materials and Structures, Institute of Innovative Research, Tokyo Institute of Technology, Yokohama 226-8503, Japan

In this part, we show the details about how we investigated the carbon defects in SiC (0001)/$SiO_2$ systems in the main body of manuscript. The material is divided into 3 sections; (i) SiC-side defects, (ii) $SiO_2$-side defects, and (iii) defects just at a SiC/$SiO_2$ interface.

(i) SiC-side defects

In investigating the carbon defects in the SiC side, we prepared a 128-atom 4H-SiC supercell (a 4×4×1 supercell) as shown in Fig. 4. We introduced 1 or 2 additional carbon atoms into this supercell into the positions described in Fig. 5, and performed structural optimization to obtain the stable defect configurations. For instance, if we introduce a C atom in the position *a* in Fig. 5, we obtain the carbon antisite, $C_{Si}$. Detail conditions of the structural optimization are described in the main body of manuscript.

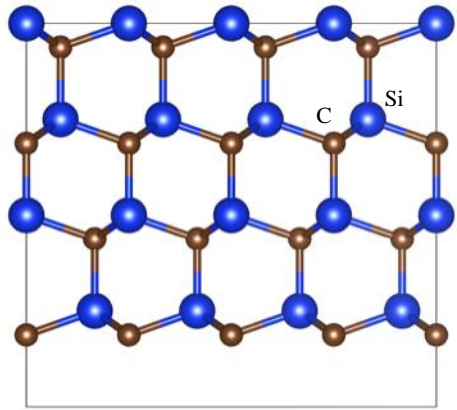

Fig. 4. A 128-atom 4H-SiC supercell (a 4×4×1 supercell) prepared in order to investigate the carbon defects in SiC.

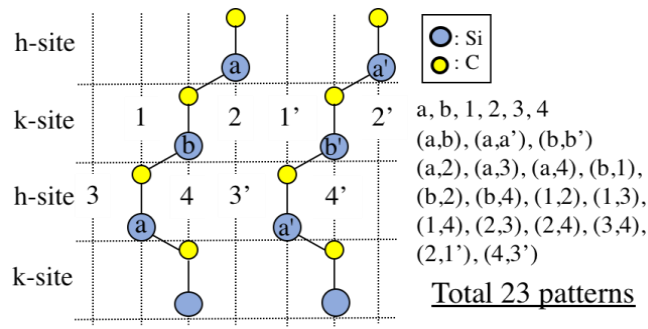

Fig. 5. Positions to introduce 1 or 2 carbon atoms in 4H-SiC (23 patterns). Position *a*, *b*, *a'*, and *b'* are the Si-sites, and position 1, 2, 3, 4, 1', 2', 3', and 4' are the interstitial sites.

(ii) SiO$_2$-side defects

In investigating the carbon defects in the SiO$_2$ side, we prepared a 72-atom α-quartz-SiO$_2$ supercell as shown in Fig. 6. We especially focused on the di-carbon defects which are reported to be stable from first-principles molecular dynamics (MD) calculations.[14] We reproduced the defect structures (Fig. 7) in the α-quartz-SiO$_2$ supercell (Fig. 6) and performed structural optimization. Detail conditions of the structural optimization are described in the main body of manuscript.

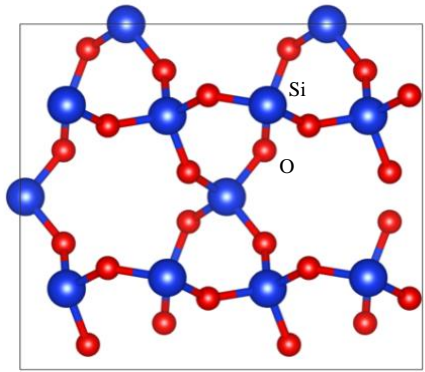

Fig. 6. A 72-atom α-quartz-SiO$_2$ supercell prepared in order to investigate the carbon defects in SiO$_2$.

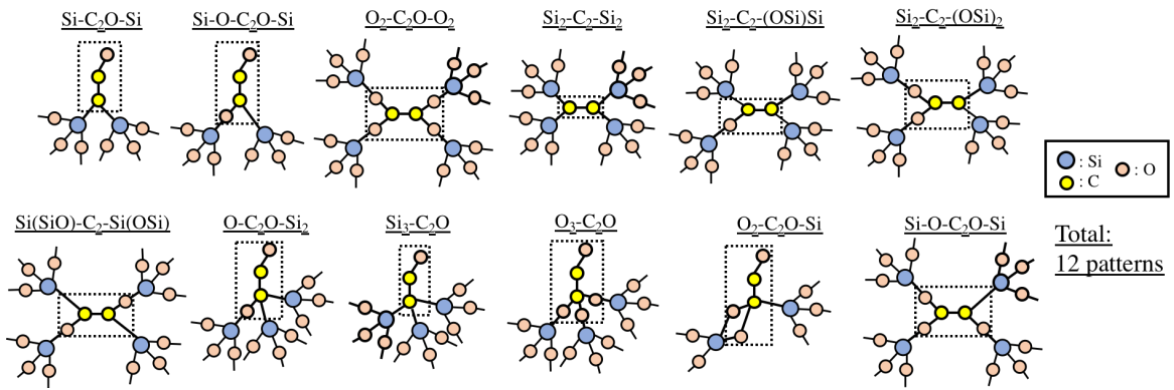

Fig. 7. Investigated structures for C-related defects in SiO$_2$. The structures were selected on the basis of first-principles molecular dynamics (MD) calculation[14] (12 patterns).

(iii) Defects just at a SiC (0001)/SiO$_2$ interface

In investigating the defects at a SiC (0001)/SiO$_2$ interface, we prepared a 128-atom 4H-SiC surface slab of which surface atoms were terminated by 32 H atoms (Fig. 8). The H atoms at a (0001) Si-face were introduced in order to eliminate the fictitious dangling bonds of surface Si atoms, which are supposed to be attached to O atoms of amorphous SiO$_2$ in the realistic SiC (0001)/SiO$_2$ structure. The surface C atoms at a (000-1) C-face should be attached to Si atoms of SiC in the realistic situation, which were also terminated by H atoms. We adopted this surface-slab model instead of preparing SiC (0001)/amorphous-SiO$_2$ structure, to eliminate the energy cost for the surrounding amorphous SiO$_2$ structure which depends on each specimen (i.e., each defect structure), and to compare the formation energy of each carbon defect on equal footing. As shown in Fig. 9, carbon atoms were inserted on the top of the surface Si atoms at a SiC (0001) surface with removing several H atoms from the surface. For example, in preparing c1d1 defect series in Fig. 9, we set a C atom above the center of gravity of 3 topmost Si atoms with removing 3 surface H atoms. Then, 3 C-Si bonds are generated. A dangling bond is left on the inserted C atom, which is treated by 4 types of termination structures (Fig. 10); a dangling bond (d), O-termination (o), OH-termination (h), and SiH$_3$-termination (s). Finally, structural optimization is performed to obtain the stable defect configurations. Detail conditions of the structural optimization are described in the main body of manuscript.

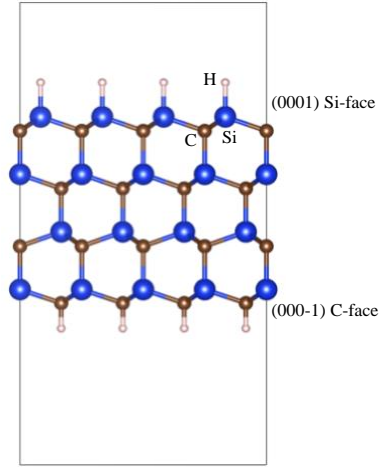

Fig. 8. A 128-atom 4H-SiC surface slab prepared in order to investigate the carbon defects at a SiC/SiO$_2$ interface. The surface Si and C atoms are terminated by 32 H atoms.

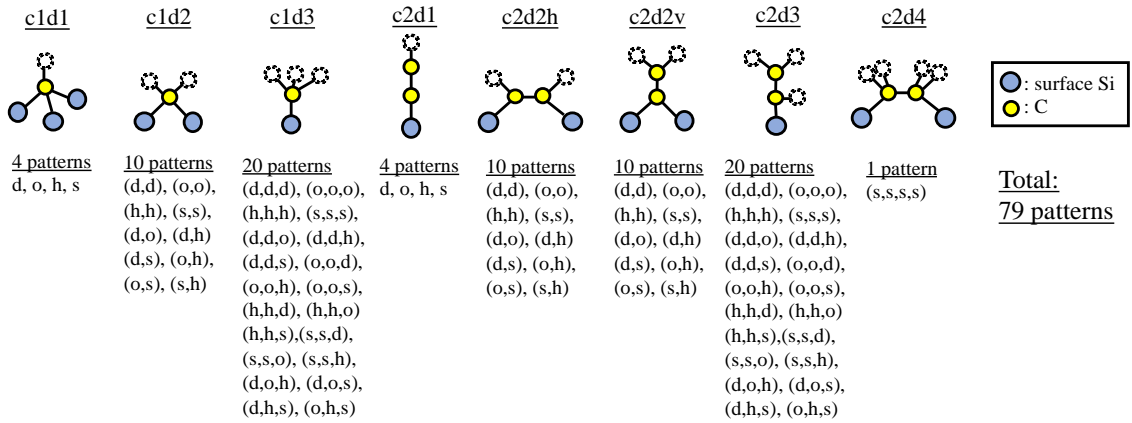

Fig. 9. Investigated structures for C-related defects at a SiC (0001)/SiO$_2$ interface. For the dangling bonds (depicted in dotted circles) of each structure, we prepared 4 types of termination structures; a dangling bond (d), O- (o), OH- (h), and SiH$_3$-termination (s), which are described in Fig. 10.

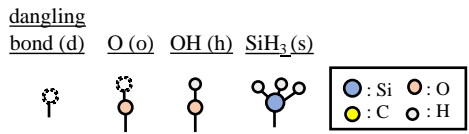

Fig. 10. 4 types of termination structures for each dangling bond of the interface carbon atoms, which is depicted as a dotted circle in Fig. 9; a dangling bond (d), O- (o), OH- (h), and SiH$_3$-termination (s). Note that the H atoms of OH- and SiH$_3$-termination are fictitious, and in the realistic situation, defects are supposed to be attached to either Si or O atom of amorphous SiO$_2$ instead of H atoms.